\documentclass[twoside,australian,sort&compress]{iopart}
\usepackage[T1]{fontenc}
\usepackage[utf8]{inputenc}
\usepackage{geometry}
\usepackage{amsbsy}
\usepackage{graphicx}
\usepackage[numbers]{natbib}

\makeatletter
\usepackage{iopams}
\usepackage{setstack}

\newcommand{\eqref}[1]{(\ref{#1})}

\usepackage{xurl}

\makeatother

\usepackage{babel}
\begin{document}
\title{Self-consistent electron-THF cross sections derived using data-driven
swarm analysis with a neural network model}
\author{{\Large{}P W Stokes$^{1}$, M J E Casey$^{1}$, D G Cocks$^{2}$,
J de Urquijo$^{3}$, G García$^{4}$, M J Brunger$^{5,6}$ and R D
White$^{1}$}}
\address{{\Large{}$^{1}$}{\large{}College of Science and Engineering, James
Cook University, Townsville, QLD 4811, Australia}}
\address{{\Large{}$^{2}$}{\large{}Research School of Physics, Australian National
University, Canberra, ACT 0200, Australia}}
\address{{\Large{}$^{3}$}{\large{}Instituto de Ciencias Físicas, Universidad
Nacional Autónoma de México, 62251, Cuernavaca, Morelos, Mexico}}
\address{{\Large{}$^{4}$}{\large{}Instituto de Física Fundamental, CSIC, Serrano
113-bis, 28006 Madrid, Spain}}
\address{{\Large{}$^{5}$}{\large{}College of Science and Engineering, Flinders
University, Bedford Park, Adelaide, SA 5042, Australia}}
\address{{\Large{}$^{6}$}{\large{}Department of Actuarial Science and Applied
Statistics, Faculty of Business and Management, UCSI University, Kuala
Lumpur 56000, Malaysia}}
\ead{{\large{}peter.stokes@my.jcu.edu.au}}
\begin{abstract}
We present a set of self-consistent cross sections for electron transport
in gaseous tetrahydrofuran (THF), that refines the set published in
our previous study \citep{DeUrquijo2019a} by proposing modifications
to the quasielastic momentum transfer, neutral dissociation, ionisation
and electron attachment cross sections. These adjustments are made
through the analysis of pulsed-Townsend swarm transport coefficients,
for electron transport in pure THF and in mixtures of THF with argon.
To automate this analysis, we employ a neural network model that is
trained to solve this inverse swarm problem for realistic cross sections
from the LXCat project. The accuracy, completeness and self-consistency
of the proposed refined THF cross section set is assessed by comparing
the analysed swarm transport coefficient measurements to those simulated
via the numerical solution of Boltzmann's equation.
\end{abstract}
\noindent{\it Keywords\/}: {swarm analysis, machine learning, artificial neural network\\
}
\submitto{\PSST }
\maketitle

\section{\label{sec:Introduction}Introduction}

Accurate modelling of electron transport through human tissue is essential
for a number of medical applications, including for treatment planning
in medical physics, and for the control and optimisation of low-temperature
atmospheric-pressure plasmas in plasma medicine \citep{Kong2009,Laroussi2009,Montie2000,Fridman2006,Nastuta2011,Miller2016}.
To accurately simulate electron transport in biological media, a precise
description of the energy deposition and electron loss/production
from scattering with each constituent biomolecule is necessary. This
description, which takes the form of electron impact cross sections
\citep{Tanaka2016}, is required over a wide range of energies, as
even subionising electrons are capable of damaging DNA through the
process of dissociative electron attachment (DEA) \citep{Boudaiffa2000,Sanche2005,Alizadeh2015}.

One of the most well-studied biomolecules, after water, is tetrahydrofuran
(THF, $\mathrm{C}_{4}\mathrm{H}_{8}\mathrm{O}$), a simple surrogate
for the complex sugar linking phosphate groups in the backbone of
DNA \citep{Brunger2017,Bug2017}. As such, numerous electron scattering
cross sections have been measured and derived for THF. These include
both experimental and theoretical derivations of the total \citep{Zecca2005,Mozejko2006a,Baek2012,Fuss2014,Bug2017},
quasielastic \citep{Colyer2007,Dampc2007,Allan2007,Homem2009,Gauf2012,Baek2012,Zhang2017},
vibrational excitation \citep{Allan2007,Dampc2007a,Khakoo2013,Do2015,Duque2015},
discrete electronic-state excitation \citep{Do2011,Zubek2011}, ionisation
\citep{Mozejko2005,Fuss2009,Dampc2011,Champion2013,Builth-Williams2013,Bug2017,Swadia2017},
and DEA cross sections \citep{Aflatooni2006,Janeckova2014}. In total,
six full sets of THF cross sections have been constructed. Chronologically,
these are due to Garland \textit{et al}. \citep{Garland2013}, for
incident electron energies from $0.1\ \mathrm{eV}$ to $300\ \mathrm{eV}$,
Fuss \textit{et al}. \citep{Fuss2014}, for energies from $1\ \mathrm{eV}$
to $10\ \mathrm{keV}$, Bug \textit{et al}. \citep{Bug2017}, for
energies from $30\ \mathrm{eV}$ to $1\ \mathrm{keV}$, Swadia \textit{et
al}. \citep{Swadia2017,Swadia2017a}, for energies from the ionisation
threshold to $5\ \mathrm{keV}$, and Casey \textit{et al}. \citep{Casey2017}
who refined the Garland \textit{et al}. set by performing and analysing
the first experimental measurements of swarm transport coefficients
in pure THF. Subsequently, de Urquijo \textit{et al}. \citep{DeUrquijo2019a}
further refined the Casey \textit{et al}. set by including transport
coefficients for admixtures of THF in argon and nitrogen in the analysis.
In the latter two studies, the \textit{inverse swarm problem} of unfolding
cross sections from swarm data was solved iteratively through the
repeated adjustment of the cross section set until a good agreement
was found between the simulated transport coefficients and experiment.

Swarm experiments provide a useful way to assess the accuracy and
self-consistency of cross sections \citep{White2018}. The iterative
approach described above for analysing swarm data dates back to Mayer
\citep{Mayer1921}, Ramsauer \citep{Ramsauer1921} and Townsend \textit{et
al.} \citep{Townsend1922}, who simulated swarm transport coefficients
for comparison with experiment using approximate forms of the electron
energy distribution function (EEDF). Since then swarm analysis has
increased in sophistication, in particular since Phelps and collaborators
\citep{Frost1962,Engelhardt1963,Engelhardt1964,Hake1967,Phelps1968}
began determining the EEDF accurately through the numerical solution
of Boltzmann's equation. Despite such improvements, it is important
to note that, as an inverse problem, swarm analysis can become ill-posed
when the amount of available experimental data is limited. That is,
multiple underlying cross section sets can potentially result in the
same collection of swarm transport coefficients. The success of iterative
swarm analysis is thus often predicated on an expert performing the
cross section adjustments, relying on their experience and intuition
in order to avoid solutions that are unphysical. This holds true even
for automated methods for swarm analysis via the numerical optimisation
of transport coefficients \citep{Duncan1972,OMalley1980,Taniguchi1987,Suzuki1989,Suzuki1990,Morgan1991a,Morgan1993,Brennan1993}
which, due to the ill-posed nature of the inverse swarm problem, can
potentially become stuck in unphysical local minima that require the
subsequent intervention and appraisal of an expert. In our recent
work \citep{Stokes2019}, we attempted to automate this expertise
by training an artificial neural network model on cross sections derived
from the LXCat project \citep{Pancheshnyi2012,Pitchford2017,LXCat}.
This neural network was applied quite successfully toward simultaneously
deriving multiple cross sections of helium from simulated swarm data,
showing the promise of this machine learning approach.

In this investigation, we apply the aforementioned data-driven swarm
analysis in order to try and determine plausible improvements to the
set of THF cross sections constructed by de Urquijo \textit{et al}.
\citep{DeUrquijo2019a}. We begin in Section \ref{sec:Neural-network-regression}
by outlining a suitable neural network for electron-THF swarm analysis,
as well as an appropriate training procedure and a suitable set of
training data. In Section \ref{sec:Machine-fitted-THF-cross}, we
apply this neural network in order to analyse pulsed-Townsend drift
velocities and Townsend first ionisation coefficients of electron
transport in both pure THF and mixtures of THF in argon. As output
from the network, we obtain for THF a quasielastic momentum transfer
cross section (MTCS), a pair of neutral dissociation cross sections,
an ionisation cross section, and an electron attachment cross section.
With these machine-fitted cross sections in place of their counterparts
in the de Urquijo \textit{et al}. set, we subsequently simulate pulsed-Townsend
transport coefficients in Section \ref{sec:Transport-coefficients-for}
to confirm that they coincide with the experimental measurements that
were used as input to the neural network. Finally, we present conclusions
in Section \ref{sec:Conclusion} while also discussing avenues for
future work.

\section{\label{sec:Neural-network-regression}Neural network for electron-THF
swarm analysis}

In this section, we provide a brief overview of the architecture and
training of our neural network for the regression of THF cross sections
given relevant sets of electron swarm transport coefficients. A more
detailed introduction to this machine-assisted approach to swarm analysis
can be found in our previous work \citep{Stokes2019}.

\subsection{\label{subsec:Architecture}Architecture}

\begin{figure}
\begin{centering}
\includegraphics[scale=0.5]{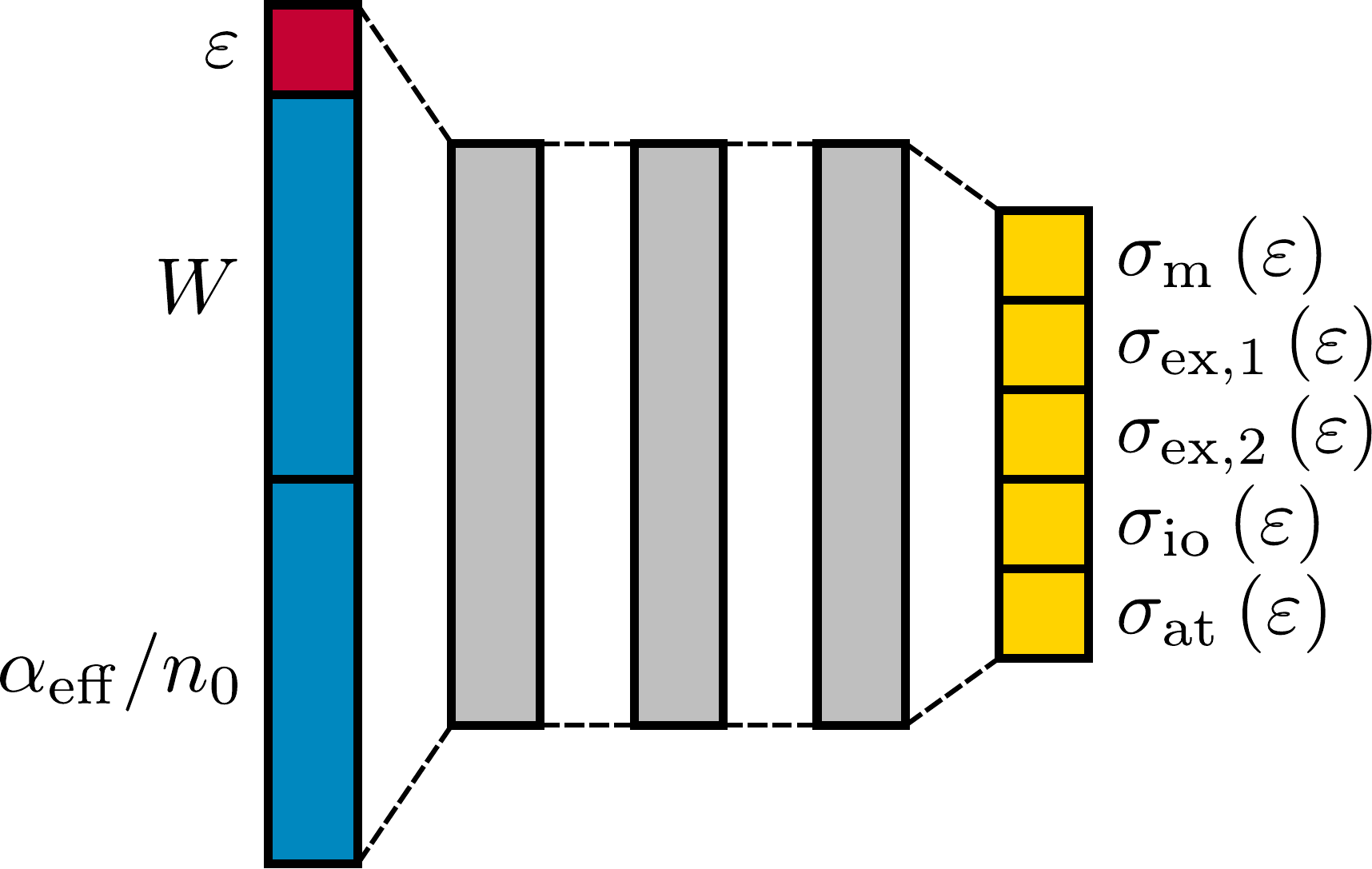}
\par\end{centering}
\caption{\label{fig:Diagram-of-the}Diagram of the fully-connected neural network,
Eq. \eqref{eq:neuralnet}, used for the regression of selected THF
cross sections (yellow) as a function of energy (red) given some relevant
electron swarm data (blue). Specifically, as output, the neural network
provides the quasielastic MTCS, $\sigma_{\mathrm{m}}\left(\varepsilon\right)$,
the neutral dissociation cross sections, $\sigma_{\mathrm{ex,1}}\left(\varepsilon\right)$
and $\sigma_{\mathrm{ex,2}}\left(\varepsilon\right)$, the ionisation
cross section $\sigma_{\mathrm{io}}\left(\varepsilon\right)$, and
the electron attachment cross section, $\sigma_{\mathrm{at}}\left(\varepsilon\right)$.
As input, in addition to the energy $\varepsilon$, the network takes
drift velocities, $W$, and effective Townsend first ionisation coefficients,
$\alpha_{\mathrm{eff}}/n_{0}$, both of which are measured for a variety
of reduced electric fields, $E/n_{0}$, and admixture ratios of THF
in argon. Cross section training data is chosen carefully, as described
in Section \ref{subsec:Cross-section-training}, so as to constrain
the derived cross sections to be within the vicinity of their known
uncertainties.}
\end{figure}
To obtain a solution to the inverse swarm problem for electron transport
in THF, we apply a fully-connected neural network in order to determine
the quasielastic (elastic+rotational) MTCS, $\sigma_{\mathrm{m}}\left(\varepsilon\right)$,
the pair of neutral dissociation cross sections, $\sigma_{\mathrm{ex,1}}\left(\varepsilon\right)$
and $\sigma_{\mathrm{ex,2}}\left(\varepsilon\right)$, the ionisation
cross section, $\sigma_{\mathrm{io}}\left(\varepsilon\right)$, and
the electron attachment cross section, $\sigma_{\mathrm{at}}\left(\varepsilon\right)$,
as illustrated by Figure \ref{fig:Diagram-of-the}. The remaining
excitation cross sections (e.g. for vibrational excitation and discrete
electronic-state excitation) are not included here, as they are considered
to be better known \citep{DeUrquijo2019a}, and are instead sourced
from the cross section set constructed by de Urquijo \textit{et al}.
\citep{DeUrquijo2019a}. The neural network performs a nonlinear mapping
from an input vector $\mathbf{x}$ containing swarm data, to an output
vector $\mathbf{y}$ containing the aforementioned cross sections:
\begin{equation}
\mathbf{y}=\left[\begin{array}{c}
\sigma_{\mathrm{m}}\left(\varepsilon\right)\\
\sigma_{\mathrm{ex,1}}\left(\varepsilon\right)\\
\sigma_{\mathrm{ex,2}}\left(\varepsilon\right)\\
\sigma_{\mathrm{io}}\left(\varepsilon\right)\\
\sigma_{\mathrm{at}}\left(\varepsilon\right)
\end{array}\right].
\end{equation}
As each output cross section is a function of energy, $\varepsilon$,
this energy is made an input to the neural network, alongside the
swarm transport coefficients:
\begin{equation}
\mathbf{x}=\left[\begin{array}{c}
\varepsilon\\
\hline W_{1}\\
W_{2}\\
\vdots\\
\hline \left(\alpha_{\mathrm{eff}}/n_{0}\right)_{1}\\
\left(\alpha_{\mathrm{eff}}/n_{0}\right)_{2}\\
\vdots
\end{array}\right],
\end{equation}
where $W$ denotes the flux drift velocity, $\alpha_{\mathrm{eff}}/n_{0}$
denotes the reduced effective Townsend first ionisation coefficient,
and $n_{0}$ is the background neutral number density. Subscripts
indicate that a number of pulsed-Townsend swarm measurements are provided
as input to the network. Mathematically, the neural network takes
the form of the following composition of functions:
\begin{equation}
\mathbf{y}\left(\mathbf{x}\right)=\left(\mathbf{A}_{4}\circ\mathrm{swish}\circ\mathbf{A}_{3}\circ\mathrm{swish}\circ\mathbf{A}_{2}\circ\mathrm{swish}\circ\mathbf{A}_{1}\right)\left(\mathbf{x}\right),\label{eq:neuralnet}
\end{equation}
where the $\mathrm{swish}\left(x\right)\equiv x/\left(1+e^{-x}\right)$
nonlinear \textit{activation function} \citep{Ramachandran2017} is
applied element-wise throughout, and each $\mathbf{A}_{n}\left(\mathbf{x}\right)\equiv\mathbf{W}_{n}\mathbf{x}+\mathbf{b}_{n}$
is an affine transformation defined by a parameter matrix $\mathbf{W}_{n}$
and vector $\mathbf{b}_{n}$. It is these parameters that are optimised
when training the neural network, as described in Section \ref{subsec:Training}
below. Note that the vectors $\mathbf{b}_{1}$, $\mathbf{b}_{2}$
and $\mathbf{b}_{3}$ are each made to contain $256$ parameters,
while $\mathbf{b}_{4}$ must contain $5$ parameters, corresponding
to the number of output cross sections. The matrices $\mathbf{W}_{n}$
are sized accordingly.

Finally, it is important to note that in what follows, the cross sections,
energies, and transport coefficients are all log-transformed before
being used to train the network, so as to ensure all training data
lies within the domain $\left[-1,1\right]$:
\begin{equation}
z\mapsto\log_{\sqrt{\frac{z_{\mathrm{max}}}{z_{\mathrm{min}}}}}\left(\frac{z}{\sqrt{z_{\mathrm{max}}z_{\mathrm{min}}}}\right),\label{eq:log-transform}
\end{equation}
where $z_{\min}$ and $z_{\mathrm{max}}$ are the extrema of all values
of the quantity $z$ employed for training. As this transformation
is undefined when $z$ is a cross section equal to zero, we replace
such instances with a suitably small positive number, which we take
to be $10^{-26}\ \mathrm{m}^{2}$. In turn, if the neural network
outputs a cross section less than $10^{-26}\ \mathrm{m}^{2}$, we
treat that output as being equal to zero instead. Threshold energies
for the processes of neutral dissociation and ionisation can thus
be inferred directly from the output of the neural network.

\subsection{\label{subsec:Cross-section-training}Cross section training data}

\begin{figure}
\begin{centering}
\includegraphics[scale=0.4]{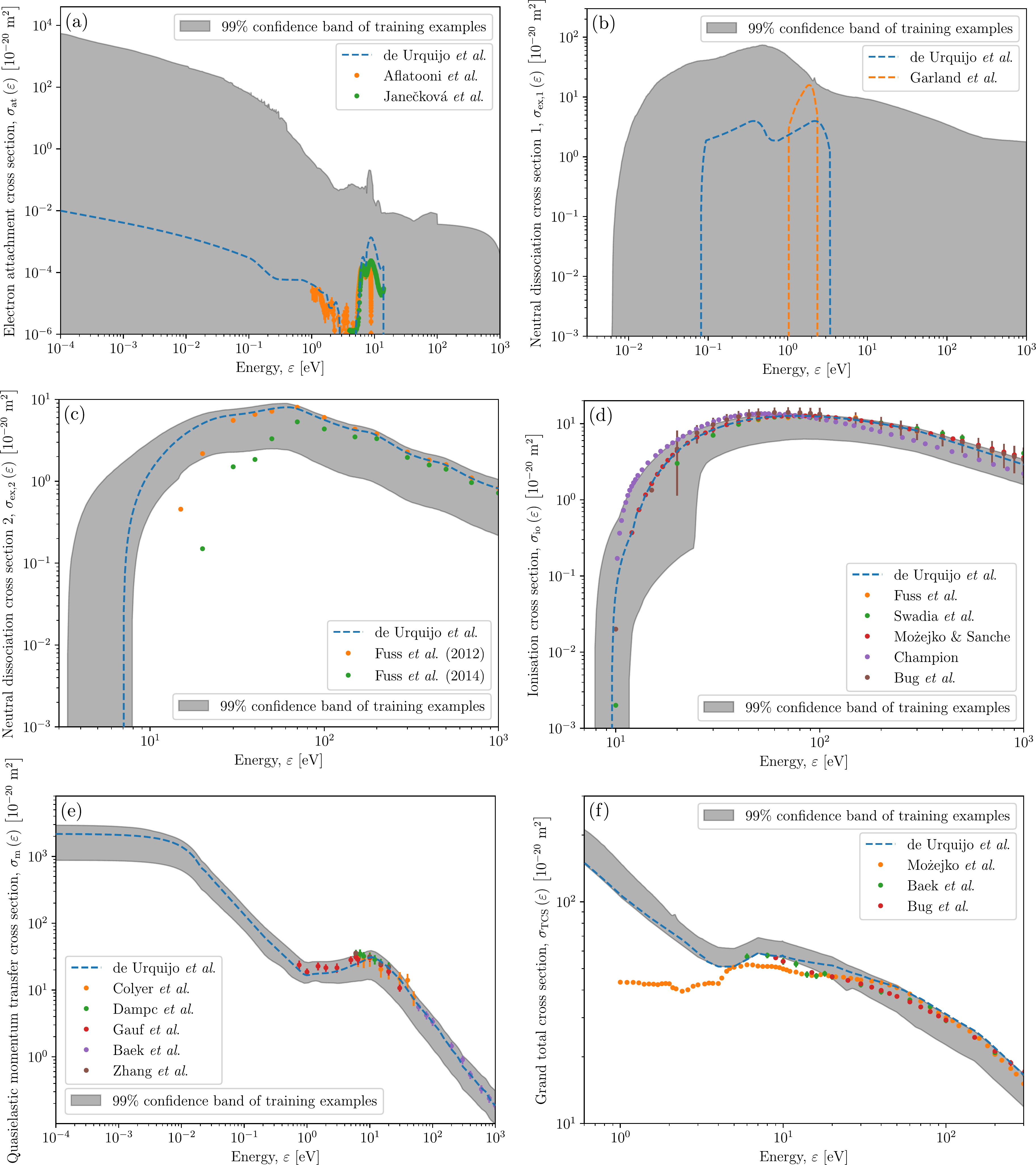}
\par\end{centering}
\caption{\label{fig:Confidence-band-of}Confidence bands (grey) for exemplar
cross sections used to train the neural network, Eq. \eqref{eq:neuralnet}.
These training cross sections are derived from the LXCat project \citep{Pancheshnyi2012,Pitchford2017,LXCat}
using Eq. \eqref{eq:mixture} and detailed in Section \ref{subsec:Cross-section-training}.
Through this choice of training data, the neural network is encouraged
to derive a cross section set that is consistent with experimental
and theoretical results from earlier studies \citep{DeUrquijo2019a,Aflatooni2006,Janeckova2014,Garland2013,Fuss2012,Fuss2014,Swadia2017,Mozejko2005,Champion2013,Bug2017,Colyer2007,Dampc2007,Gauf2012,Baek2012,Zhang2017,Mozejko2006},
including those from the recent set of de Urquijo \textit{et al.}
\citep{DeUrquijo2019a} (blue dashed lines).}
\end{figure}
We construct exemplar cross sections for training the neural network,
Eq. \eqref{eq:neuralnet}, through the pairwise geometric combination
of cross sections from the LXCat project \citep{Pancheshnyi2012,Pitchford2017,LXCat,Biagi,Biagiv71,Bordage,BSR,CCC,Christophorou,COP,eMolLeHavre,FLINDERS,Hayashi,ISTLisbon,Itikawa,Morgan,NGFSRDW,Phelps,QUANTEMOL,SIGLO,TRINITI}.
That is, given a random pair of LXCat cross sections, $\sigma_{1}\left(\varepsilon\right)$
and $\sigma_{2}\left(\varepsilon\right)$, of a given type (e.g. electron
attachment, ionisation, etc.), as well as a uniformly sampled mixing
ratio $r\in\left[0,1\right]$, a physically-plausible cross section
of the same type is formed as:
\begin{equation}
\sigma\left(\varepsilon\right)=\sigma_{1}^{1-r}\left(\varepsilon+\varepsilon_{1}-\varepsilon_{1}^{1-r}\varepsilon_{2}^{r}\right)\sigma_{2}^{r}\left(\varepsilon+\varepsilon_{2}-\varepsilon_{1}^{1-r}\varepsilon_{2}^{r}\right),\label{eq:mixture}
\end{equation}
where $\varepsilon_{1}$ and $\varepsilon_{2}$ are the respective
threshold energies of $\sigma_{1}\left(\varepsilon\right)$ and $\sigma_{2}\left(\varepsilon\right)$.
This formula has the benefit of retaining the correlation between
the magnitude of a cross section and its threshold energy \citep{Stokes2019}.

We apply Eq. \eqref{eq:mixture} directly to generate suitable training
examples for the electron attachment cross section, $\sigma_{\mathrm{at}}\left(\varepsilon\right)$,
and the lower-threshold neutral dissociation cross section, $\sigma_{\mathrm{ex},1}\left(\varepsilon\right)$.
No explicit constraints are placed on these cross sections, as seen
by the large confidence bands for the training examples in Figures
\ref{fig:Confidence-band-of}(a) and (b). To emphasise this point,
although we refer to $\sigma_{\mathrm{ex},1}\left(\varepsilon\right)$
as the neutral dissociation cross section of ``lower threshold'',
some of its training examples have threshold energies that exceed
that of the ``higher threshold'' neutral dissociation cross section,
$\sigma_{\mathrm{ex},2}\left(\varepsilon\right)$.

For the remaining cross sections of interest, we choose to explicitly
constrain the training cross sections to lie within the vicinity of
the known experimental error bars so as to encourage the neural network
to also restrict its output in the same way. To do this, in each case
we apply Eq. \eqref{eq:mixture} to first generate an unconstrained
cross section by mixing the relevant LXCat cross sections, and then
we apply Eq. \eqref{eq:mixture} once more to mix this unconstrained
cross section with its counterpart from the de Urquijo \textit{et
al}. \citep{DeUrquijo2019a} set, weighting heavily toward the latter
with a mixing ratio of $r=0.9$. In this way our training cross sections
are thus energy-dependent perturbations of the respective de Urquijo
\textit{et al}. cross sections. The resulting confidence bands of
these training examples can be seen plotted in Figures \ref{fig:Confidence-band-of}(c)–(e).

Once the separate training cross sections are generated as described
above, each are used to replace their counterpart in the de Urquijo
\textit{et al}. set in order to obtain a proposed full data set of
cross sections for training. Rejection sampling is then used to only
keep generated cross section sets that have a grand total cross section
(TCS) that lies within $30\%$ of that of Fuss \textit{et al}. \citep{Fuss2014}
and Fuss \textit{et al}. \citep{Fuss2012}. This constraint is illustrated
by the confidence band in Figure \ref{fig:Confidence-band-of}(f).
In total, $5\times10^{4}$ such cross section sets are generated for
use in the the training procedure.

Note that, when training the neural network, cross sections must be
sampled at discrete points within the energy domain, which we choose
to be $\varepsilon\in\left[10^{-4}\ \mathrm{eV},10^{3}\ \mathrm{eV}\right]$.
We select such points using:
\begin{equation}
\varepsilon=10^{s}\ \mathrm{eV},\label{eq:sample}
\end{equation}
where $s\in\left[-4,3\right]$ is a uniformly distributed random number.

\subsection{\label{subsec:Transport-coefficient-training}Transport coefficient
training data}

Finally, to complete each input/output training pair, corresponding
pulsed-Townsend swarm transport coefficients must be simulated. For
this, we apply the two-term approximation \citep{Hagelaar2005,Robson1986}
to Boltzmann's equation and then perform backward prolongation \citep{Sherman1960}
of the EEDF by inward integration from high to low energies, using
an adaptive order adaptive energy Adams-Moulton method \citep{Hairer1993},
as implemented in the \textit{DifferentialEquations.jl} software ecosystem
\citep{Rackauckas2017,DelayDiffEq,OrdinaryDiffEq}.

As input to the neural network, we use drift velocities and reduced
effective Townsend first ionisation coefficients measured using the
pulsed-Townsend technique by de Urquijo \textit{et al}. \citep{DeUrquijo2019a}
for electron transport in both pure THF, as well as in admixtures
of THF in argon. Specifically, these measurements were taken for THF
mixture ratios of $1\%$, $2\%$, $5\%$, $10\%$, $20\%$, $50\%$
and $100\%$, across a variety of reduced electric fields, $E/n_{0}$,
ranging from $0.23\ \mathrm{Td}$ to $1000\ \mathrm{Td}$, where $1\ \mathrm{Td}=1\ \mathrm{Townsend}=10^{-21}\ \mathrm{V\ m^{2}}$.
For calculating the admixture transport coefficients, we use the argon
cross section set present in the Biagi v7.1 database \citep{Biagiv71}.

To account for the random error present in experimental measurements,
we augment the aforementioned simulated transport coefficients by
multiplying with a small amount of random noise before training sampled
from a log-normal distribution. To be specific, we sample the natural
logarithm of this noise factor from a normal distribution with a mean
of $0$ and a standard deviation of $0.01$.

It should be noted that we have recently come to the view that as
the experimental effective Townsend first ionisation coefficients
below $10^{-24}\ \mathrm{m^{2}}$ are at the limit of the apparatus
measurement capability, they should not be included in the present
analysis and nor should they have been considered in the analysis
of de Urquijo \textit{et al}. \citep{DeUrquijo2019a}. Because of
this, as well as discrepancies attributed to Penning ionisation \citep{DeUrquijo2019a},
we choose to exclude all $1\%$ and $2\%$ THF admixture effective
Townsend first ionisation coefficients from our analysis.

\subsection{\label{subsec:Training}Training procedure}

We implement the neural network, Eq. \eqref{eq:neuralnet}, using
the \textit{Flux.jl} machine learning framework \citep{Innes2018}.
We initialise the neural network parameters in $\mathbf{b}_{n}$ to
zero and those in $\mathbf{W}_{n}$ to uniform random numbers as described
by Glorot and Bengio \citep{Glorot2010}. Then, we use the Adam optimiser
\citep{Kingma2015}, with step size $\alpha=10^{-3}$, exponential
decay rates $\beta_{1}=0.9$ and $\beta_{2}=0.999$, and small parameter
$\epsilon=10^{-8}$, to adjust the parameters so as to minimise the
mean absolute error of the cross sections fitted by the neural network:
\begin{equation}
\frac{1}{5N}\sum_{i=1}^{N}\left\Vert \mathbf{y}_{i}-\boldsymbol{\sigma}\left(\mathbf{x}_{i}\right)\right\Vert _{1},\label{eq:MAE}
\end{equation}
where the index $i$ ranges over the entire set of $N$ training examples
$\left(\mathbf{x}_{i},\mathbf{y}_{i}\right)$, and $\boldsymbol{\sigma}\left(\mathbf{x}_{i}\right)$
is the associated neural network cross section prediction. We choose
to optimise the mean absolute error, instead of the mean squared error,
due to its robustness in the presence of outliers in the training
data, which are expected in the parts of the underlying cross sections
that are most uncertain. Specifically, the neural network parameters
are updated by the optimiser repeatedly using batches of $4096$ input/output
training examples, each consisting of $16$ random LXCat-derived cross
section sets selected from the $5\times10^{4}$ generated in total,
where each set is sampled with Eq. \eqref{eq:sample} at $256$ random
energies within the domain $\left[10^{-4}\ \mathrm{eV},10^{3}\ \mathrm{eV}\right]$.
Training is continued until the transport coefficients, resulting
from the fitted cross section set, best match the pulsed-Townsend
transport coefficients that were used to perform the fit.

\section{\label{sec:Machine-fitted-THF-cross}Machine-fitted THF cross sections}

In this section, we present the resulting electron-THF cross sections
that were determined automatically from swarm data by using the neural
network, Eq. \eqref{eq:neuralnet}, described in the previous section.
It should be noted that, when these cross sections are used to simulate/reproduce
the aforementioned swarm data, the resulting mean electron energies
for the swarms vary between $0.03\mathrm{eV}$ and $7.77\mathrm{eV}$.
As such, cross sections that are significantly outside of this energy
range are unlikely to have a large effect on the considered swarm
transport coefficients. In these regimes of very small or very large
energies, it is thus expected that the neural network would rely more
heavily on its prior knowledge of what constitutes a physically-plausible
cross section than on the swarm measurements themselves.

\subsection{Quasielastic momentum transfer cross section}

\begin{figure}
\begin{centering}
\includegraphics[scale=0.5]{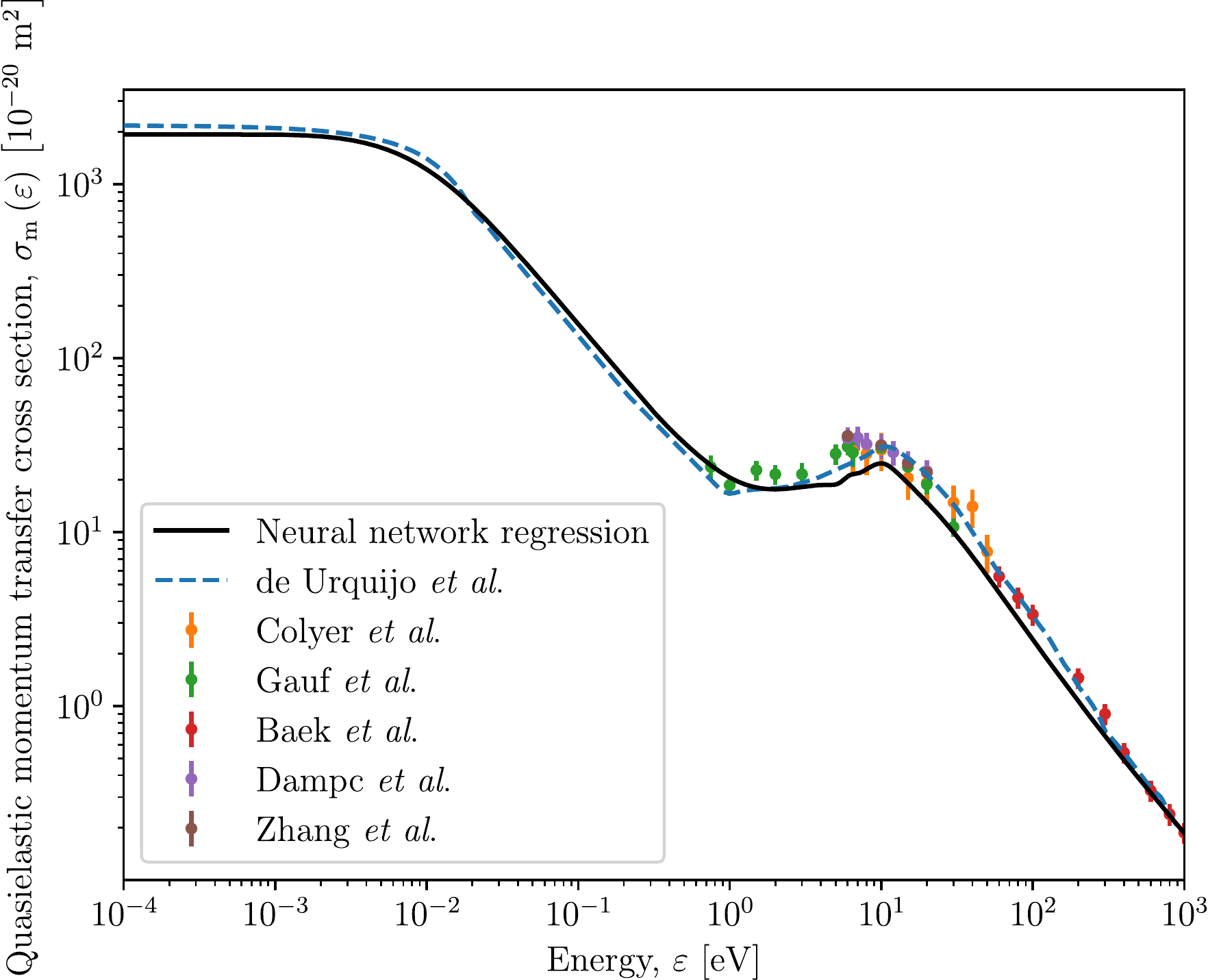}
\par\end{centering}
\caption{\label{fig:Quasielastic-momentum-transfer}Previous quasielastic momentum
transfer cross sections \citep{DeUrquijo2019a,Colyer2007,Gauf2012,Baek2012,Dampc2007,Zhang2017},
compared to that determined from our neural network regression approach.}
\end{figure}
Overall, the machine-fitted quasielastic MTCS does not deviate far
from that of de Urquijo \textit{et al}. \citep{DeUrquijo2019a}, as
shown in Figure \ref{fig:Quasielastic-momentum-transfer}, and as
such agrees reasonably well with the experimental and calculated cross
sections of Coyler \textit{et al}. \citep{Colyer2007}, Gauf \textit{et
al}. \citep{Gauf2012}, Baek \textit{et al}. \citep{Baek2012}, Dampc
\textit{et al}. \citep{Dampc2007}, and Zhang \textit{et al}. \citep{Zhang2017}.
At very low energies, below $10^{-2}\ \mathrm{eV}$, the neural network
predicts a roughly constant quasielastic MTCS that is about $10\%$
smaller in magnitude compared to that of the de Urquijo \textit{et
al}. counterpart in the same energy regime. The greatest relative
deviation from the de Urquijo \textit{et al}. cross section occurs
around $25\ \mathrm{eV}$, where the cross section determined by the
neural network is smaller by $30\%$.

\subsection{Neutral dissociation cross section}

\begin{figure}
\begin{centering}
\includegraphics[scale=0.5]{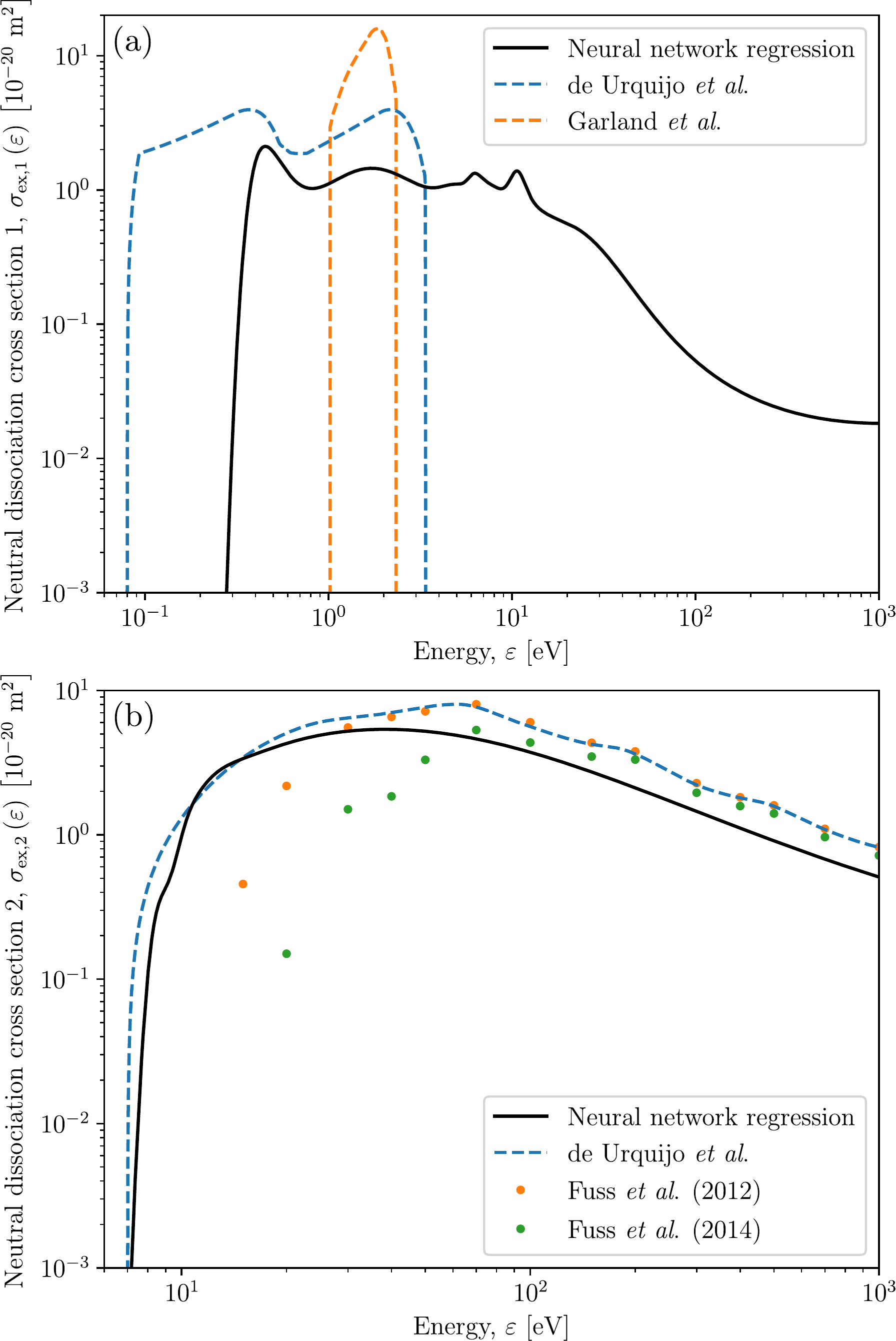}
\par\end{centering}
\caption{\label{fig:Neutral-dissociation-cross}A comparison of the low and
high energy neutral dissociation cross sections, (a) and (b), as determined
from the present neural network regression and from previous studies
\citep{DeUrquijo2019a,Garland2013,Fuss2012,Fuss2014}.}
\end{figure}

The low-threshold energy neutral dissociation cross section found
by the neural network highlights the non-uniqueness of this inverse
swarm problem, as it differs substantially from that of both Garland
\textit{et al}. \citep{Garland2013} and de Urquijo \textit{et al}.
\citep{DeUrquijo2019a}, as seen in Figure \ref{fig:Neutral-dissociation-cross}(a).
To begin with, the fitted threshold energy is equal to $0.23\ \mathrm{eV}$,
lying between the thresholds of $0.08\ \mathrm{eV}$ for de Urquijo
\textit{et al}. and $1\ \mathrm{eV}$ for Garland \textit{et al}.
Additionally, the cross section magnitude is also smaller than both
aforementioned counterparts, with a peak of $2\times10^{-20}\ \mathrm{m^{2}}$.
From its maximum value, this neutral dissociation cross section remains
roughly constant until $10\ \mathrm{eV}$, where it decays by roughly
two orders of magnitude by $1000\ \mathrm{eV}$.

The fitted high-threshold energy neutral dissociation cross section,
plotted in Figure \ref{fig:Neutral-dissociation-cross}(b), can be
seen to have a smaller threshold energy of $6.3\ \mathrm{eV}$ compared
to the $7\ \mathrm{eV}$ used by de Urquijo \textit{et al}. In general,
this cross section prediction lies below its de Urquijo \textit{et
al}. counterpart, by up to $40\%$ at high energies. This puts this
machine-fitted cross section more in line at higher energies with
the results of Fuss \textit{et al}. \citep{Fuss2014}, compared to
those of Fuss \textit{et al}. \citep{Fuss2012}.

\subsection{Ionisation cross section}

\begin{figure}
\begin{centering}
\includegraphics[scale=0.5]{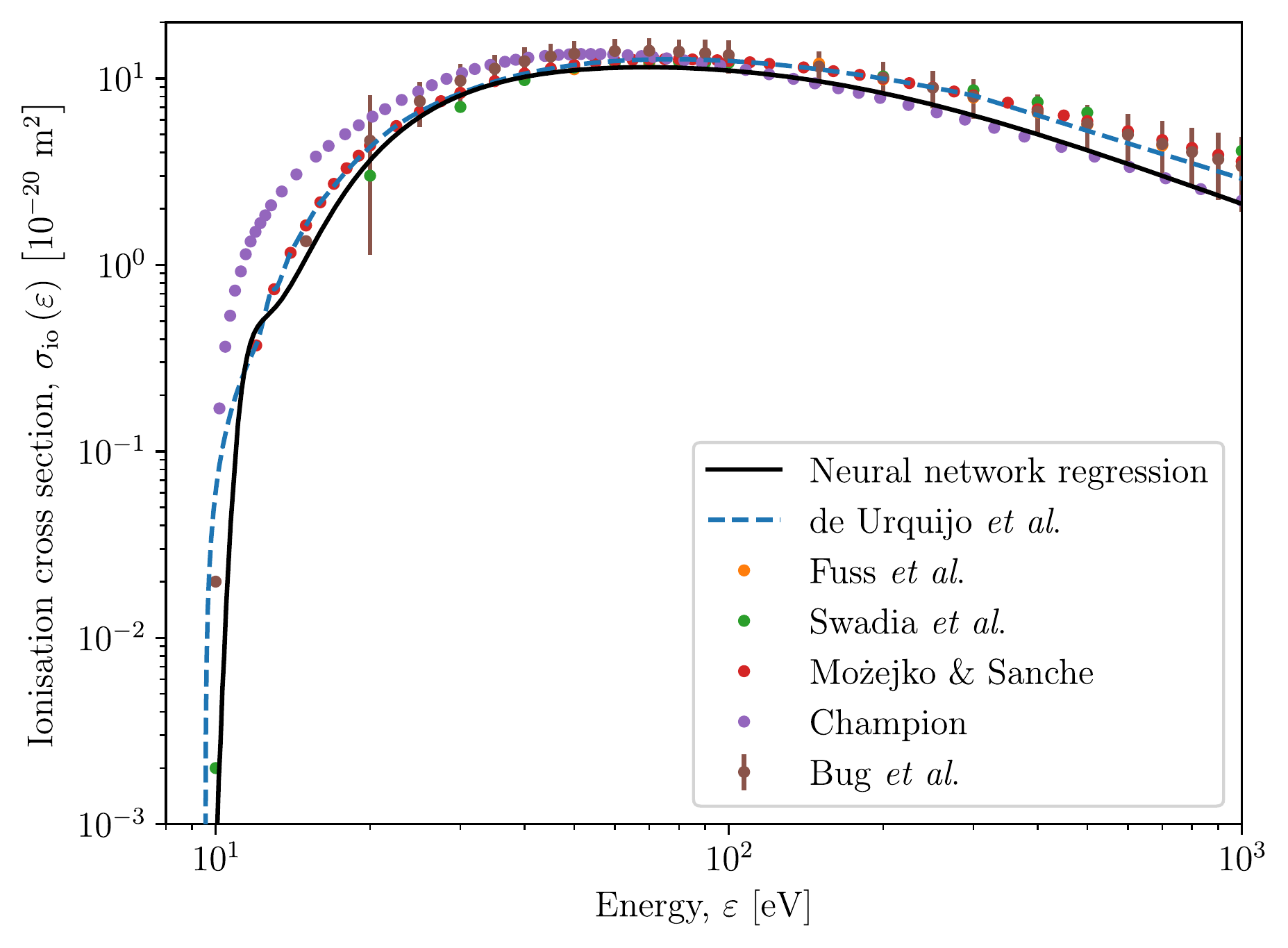}
\par\end{centering}
\caption{\label{fig:Ionisation-cross-section}A comparison of the present neural
network regression ionisation cross section, with a selection of earlier
results \citep{DeUrquijo2019a,Fuss2014,Swadia2017,Mozejko2005,Champion2013,Bug2017}.}
\end{figure}

The neural network prediction for the ionisation cross section, plotted
in Figure \ref{fig:Ionisation-cross-section}, is agrees fairly well
that of the de Urquijo \textit{et al}. \citep{DeUrquijo2019a} at
low to intermediate energies, up to $100\ \mathrm{eV}$, and thus
also coincides well with the cross sections of Fuss \textit{et al}.
\citep{Fuss2014}, Swadia \textit{et al}. \citep{Swadia2017}, Możejko
and Sanche \citep{Mozejko2005}, and Bug \textit{et al}. \citep{Bug2017}.
Beyond $100\ \mathrm{eV}$, the machine-fitted cross section agrees
particularly well with the theoretical result of Champion \citep{Champion2013}.
Although the neural network regression here suggested an ionisation
threshold energy of $8.99\ \mathrm{eV}$, it should be noted that
increasing this threshold to $9.55\ \mathrm{eV}$, the value adopted
in Refs. \citep{Garland2013,Casey2017,DeUrquijo2019a} from the experimental
value of Dampc \textit{et al}. \citep{Dampc2011}, did not result
in any perceptible change to the simulated swarm transport coefficients.

\subsection{Non-dissociative/dissociative electron attachment cross section}

\begin{figure}
\begin{centering}
\includegraphics[scale=0.5]{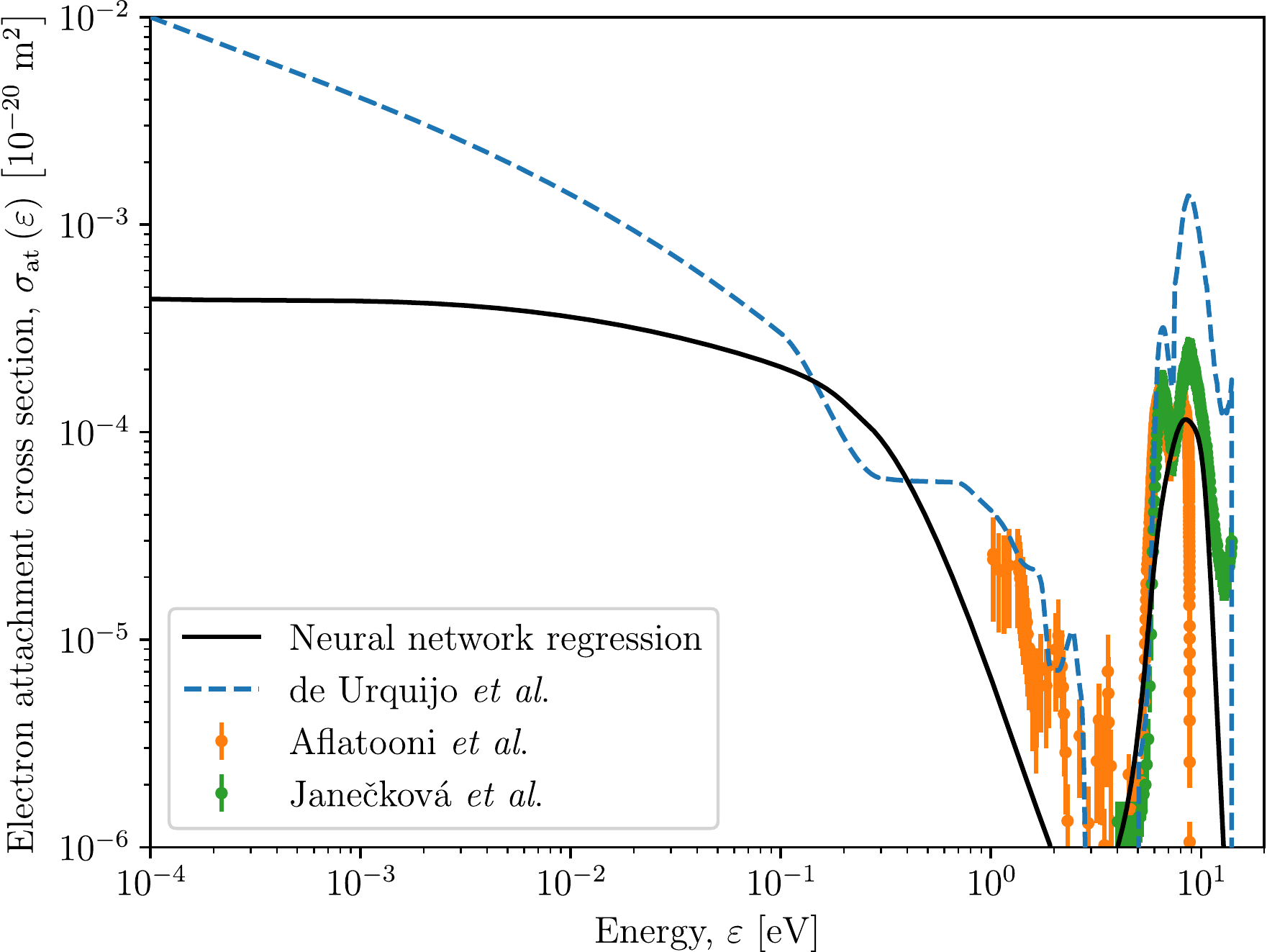}
\par\end{centering}
\caption{\label{fig:Dissociative-electron-attachment}A comparison of the present
neural network regression of the non-dissociative/dissociative electron
attachment cross section, alongside a selection of earlier results
\citep{DeUrquijo2019a,Aflatooni2006,Janeckova2014}.}
\end{figure}

The measurement of electron attachment cross sections in THF has been
concentrated mostly on dissociative electron attachment (DEA). The
only experiment reporting the direct detection of a metastable negative
ion $\left[\mathrm{THF}\right]^{*-}$ is that of Sulzer \textit{et
al}. \citep{Sulzer2006}, arising from a coordinated research between
two laboratories at Innsbruck and Berlin, which differ only in the
ion source. In the Innsbruck apparatus, the electron beam was produced
by an electrostatic hemispherical electron monochromator while at
Berlin the beam was generated from a trochoidal electron monochromator.
Both ion sources had a similar energy resolution in the range 100-130
eV. Apart from these differences, in both devices the electron beam
is made to intersect orthogonally with the effusive molecular beam.
The ions are extracted by a small electric field towards the entrance
of a quadrupole mass spectrometer and detected at its exit. A perfect
agreement between the data obtained from both laboratories was reported.

Ibănescu \textit{et al}. \citep{Ibanescu2008} used a magnetically
collimated trochoidal electron monochromator with a resolution of
about 150 meV. The beam was focused into the collision chamber filled
with THF. The collision fragment anions were extracted and focused
into a quadrupole mass spectrometer. No $\mathrm{THF^{-*}}$ was detected,
but it was recognised that the detection of these anions is very rare
in DEA experiments well above non-thermal energies. Regarding the
dissociation products, Ibănescu \textit{et al}. found that the most
abundant ion was $\mathrm{C_{2}HO^{-}}$, followed by $\mathrm{H^{-}}$
and $\mathrm{C_{2}H_{2}O^{-}}$ over the combined energy range of
5-13 eV.

Using the Innsbruck apparatus, Sulzer \textit{et al}. \citep{Sulzer2006}
detected $\mathrm{C_{4}H_{8}O^{-}}$ ($\mathrm{THF^{-}}$), $\mathrm{C_{4}H_{6}O^{-}}$
and $\mathrm{C_{2}HO^{-}}$, with the negative ion yield of $\mathrm{THF^{-}}$
peaking at about 1 eV. Provided that this anion was observed at an
energy above 1 eV, Sulzer \textit{et al}. concluded, without providing
any further explanation, that the $\mathrm{THF^{-}}$ species detected
was generated via secondary processes. In connection with this, if
the ion is formed in the collision cell, the reaction rates leading
to a secondary negative ion with a mass equal to that of THF would
have to be very high. 

Finally, Aflatooni \textit{et al}. \citep{Aflatooni2006} used a modified
electron transmission spectrometer with a resolution similar to the
above experiments and were able to measure an absolute DEA cross section
in THF over the range 1-8.6 eV.

Even though Sulzer \textit{et al}. were the only group which detected
$\mathrm{THF^{-}}$, and provided that their energy resolution hindered
them to explore lower energies close to thermal, we conclude that
the existence of a $\mathrm{THF^{-}}$ species formed by resonant
electron attachment cannot be ruled out at once. Furthermore, looking
at the $\alpha_{\mathrm{eff}}$ curves plotted in Figure \ref{fig:Corresponding-transport-coeffici}(b)
, the increasingly negative value of this swarm coefficient with decreasing
$E/n_{0}$ (i.e. mean energy) strongly suggests the possibility of
a resonantly formed $\mathrm{THF^{-}}$ species at energies well below
1 eV. In view of the need to extend the attachment cross section set
down to energies so low as $10^{-4}\ \mathrm{eV}$ in this research
where only anions from the parent molecule may form, we shall refer
to, regardless of the ion species, the present electron total attachment
cross section as the non-dissociative/dissociative attachment (NDA-DEA)
cross section. 

The NDA-DEA cross section determined by the neural network is plotted
in Figure \ref{fig:Dissociative-electron-attachment}. Below $0.1\ \mathrm{eV}$
the present neural network prediction flattens, becoming constant
in magnitude below $10^{-2}\ \mathrm{eV}$ and differing significantly
from the ``hand-fitted'' NDA-DEA proposed by de Urquijo \textit{et
al}. \citep{DeUrquijo2019a} which, by contrast, increases by over
an order of magnitude down to $10^{-4}\ \mathrm{eV}$ according to
a rough power law. Although no explicit constraints were placed on
the NDA-DEA fit — see Figure \ref{fig:Confidence-band-of}(a) for
the range of attachment training data used — the resulting neural
network regression can be seen to agree fairly well overall within
the experimental uncertainties of the measurements of both Aflatooni
\textit{et al}. \citep{Aflatooni2006} and Janečková \textit{et al}.
\citep{Janeckova2014}. Beyond $16\ \mathrm{eV}$, the neural network
does not find any noticeable DEA, even though this possibility is
by no means ruled out given the scope of examples used to train the
network.

\subsection{Grand total cross section}

\begin{figure}
\begin{centering}
\includegraphics[scale=0.5]{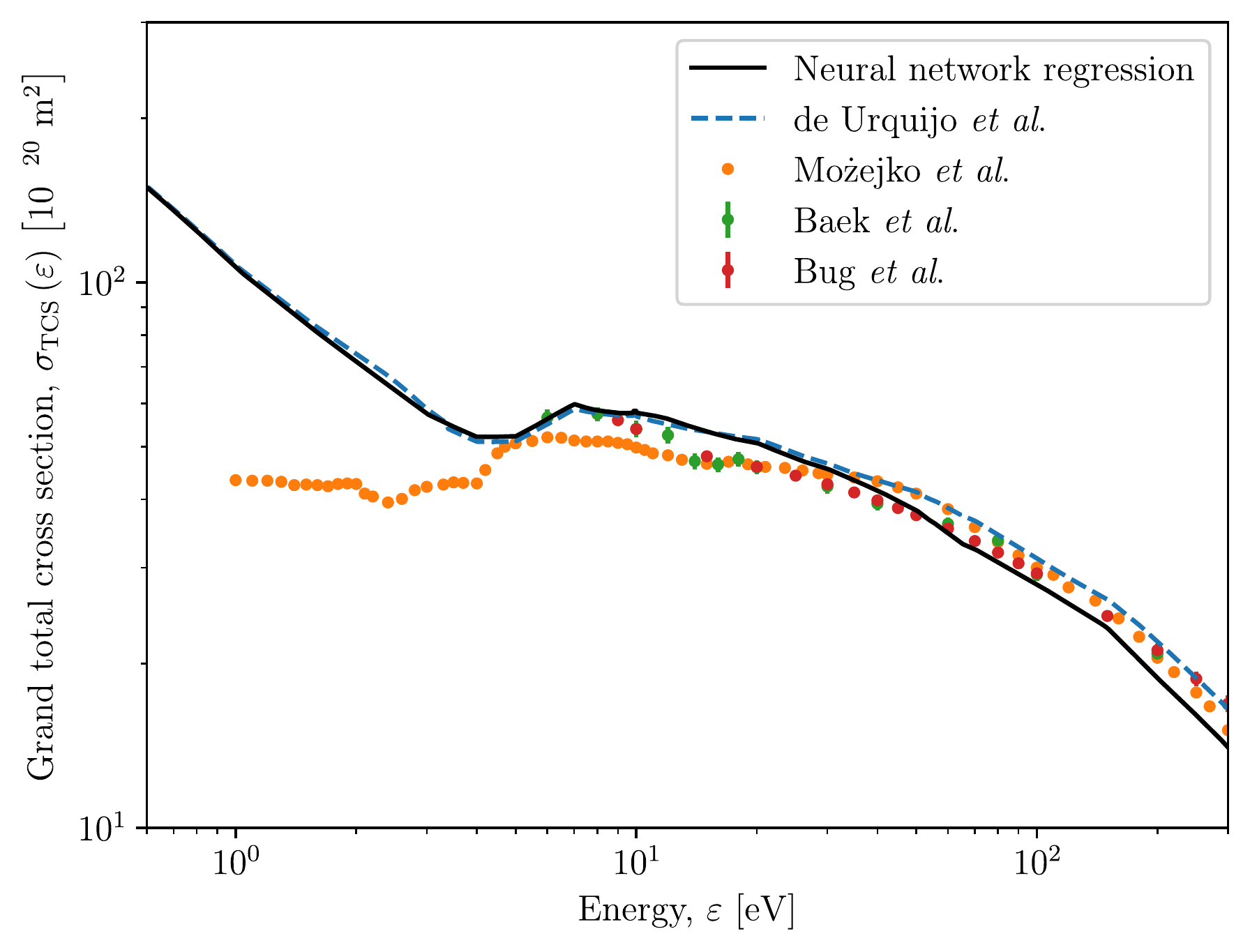}
\par\end{centering}
\caption{\label{fig:Grand-total-cross}A comparison of the present neural network
regression grand total cross section, with a selection of earlier
results \citep{DeUrquijo2019a,Mozejko2006,Baek2012,Bug2017}. }
\end{figure}
As expected from the constraints placed on the training data, the
cross sections determined by the neural network are consistent with
the grand total cross section (TCS) of the de Urquijo \textit{et al}.
\citep{DeUrquijo2019a} set. We show this in Figure \ref{fig:Grand-total-cross},
by simply summing the entire cross section set with the quasi-elastic
momentum transfer cross section replaced by the quasi-elastic integral
cross section derived by Casey \textit{et al}. \citep{Casey2017}
from the grand-total cross section of Fuss \textit{et al}. \citep{Fuss2014}.
Consequently, the resulting TCS also agrees fairly well with the experimental
measurements of Baek \textit{et al}. \citep{Baek2012}, Bug \textit{et
al}. \citep{Bug2017}, and Możejko \textit{et al}. \citep{Mozejko2006}
above $4\ \mathrm{eV}$.

\section{\label{sec:Transport-coefficients-for}Transport coefficients for
the refined cross section set}

\begin{figure}
\begin{centering}
\includegraphics[scale=0.4]{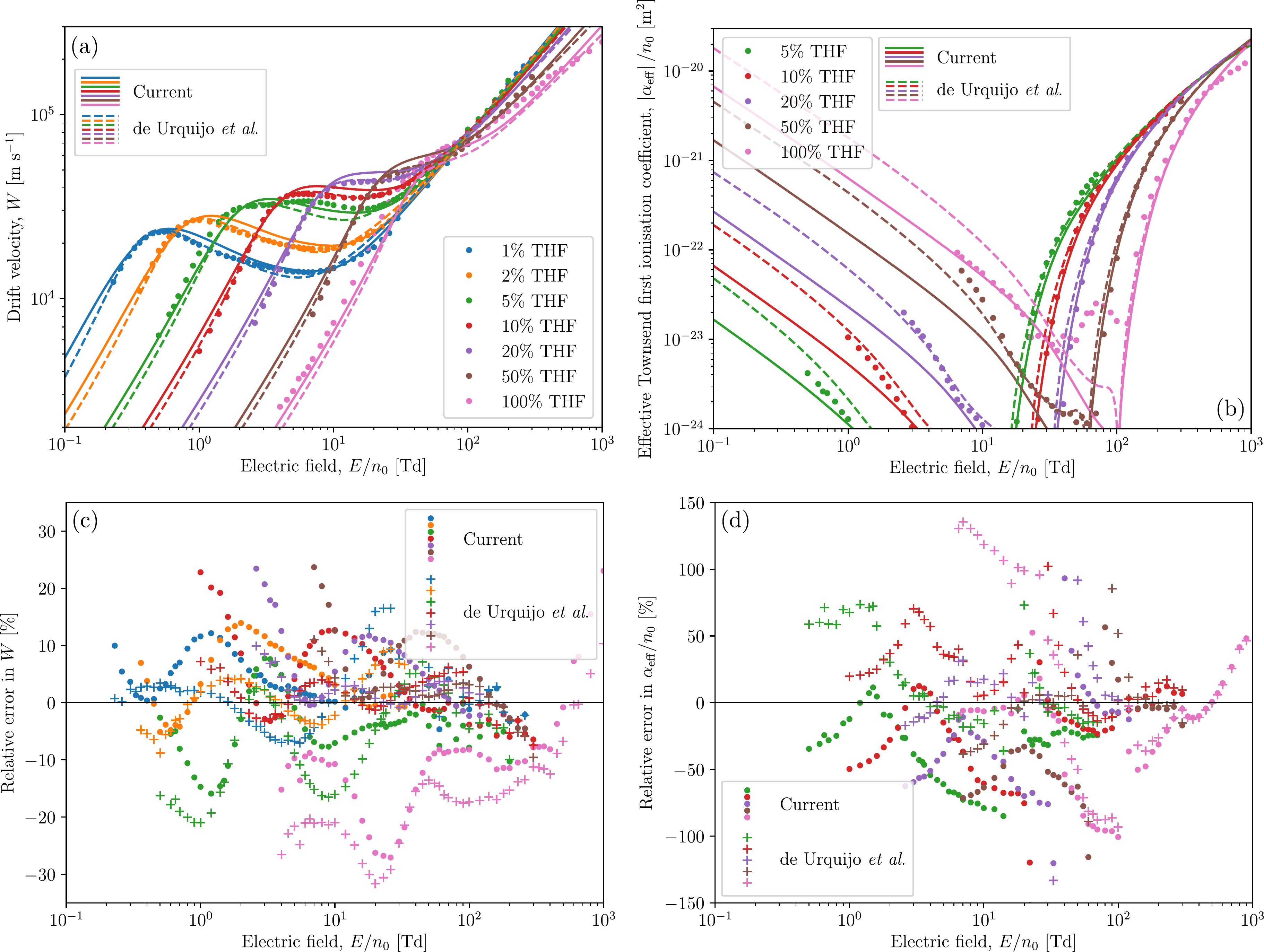}
\par\end{centering}
\caption{\label{fig:Corresponding-transport-coeffici}Simulated transport coefficients
(solid curves) of, (a), flux drift velocity, $W$, and, (b), effective
Townsend first ionisation coefficient, $\alpha_{\mathrm{eff}}/n_{0}$,
for the neural network refined cross section set, alongside corresponding
percentage error plots (c) and (d), respectively. Note that some outlying
percentage error markers in (d) have been truncated. }
\end{figure}
Transport coefficients are calculated using a two-term Boltzmann equation
solver with the machine-fitted cross sections presented in the previous
section, and are plotted in Figure \ref{fig:Corresponding-transport-coeffici}
for comparison against the measured pulsed-Townsend swarm data used
to perform the fit, as well as corresponding transport coefficient
values from the cross section data from de Urquijo \textit{et al}.
\citep{DeUrquijo2019a}. Figure \ref{fig:Corresponding-transport-coeffici}(a)
plots the drift velocities, $W$, while Figure \ref{fig:Corresponding-transport-coeffici}(b)
plots the effective Townsend first ionisation coefficients, $\alpha_{\mathrm{eff}}/n_{0}$.
In addition, Figures \ref{fig:Corresponding-transport-coeffici}(c)
and (d), respectively, plot their percentage differences relative
to the experimental swarm data.

The neural network refined cross section set can be seen to improve
the accuracy of the pure THF drift velocities, particularly at lower
reduced fields where the error is now $<5\%$. At higher fields, there
is still an improvement with the difference now $<20\%$, rather than
the $<32\%$ difference found using the cross section set of Ref.
\citep{DeUrquijo2019a}. A similar improvement can be seen for the
$5\%$ THF mixture ratio, but the same cannot be said for the remaining
THF mixture ratios which have somewhat worsened the agreement at lower
fields, possibly as a trade-off for the increased accuracy in the
$5\%$ and pure THF cases. The $10\%$, $20\%$ and $50\%$ THF mixtures
were the worst affected, with differences compared to the measured
swarm data reaching as high as $43\%$ at the lowest fields considered.

For the effective Townsend first ionisation coefficient, the modified
cross section set is seen to be generally comparable to the de Urquijo
\textit{et al}. set, at least in terms of relative error. The qualitative
form of the resultant transport coefficients, in the electronegative
region, are however, generally poorer for the modified set compared
to that for the de Urquijo \textit{et al}. \citep{DeUrquijo2019a}
set, with the exception of the case of pure THF. In the electropositive
region, the modified set results in Townsend coefficients that generally
underestimate the experimental measurements. That said, the discrepancy
between the measured and calculated effective Townsend coefficients,
in this region, has clearly improved for the $20\%$ and $50\%$ THF
mixtures, although worsened for pure THF.

Overall, we can conclude that the neural network model has produced
a plausible THF cross section set that is of comparable quality to
the recent hand-refined set of de Urquijo \textit{et al}. \citep{DeUrquijo2019a},
while importantly being free from the subjectivity inherent to conventional
swarm analysis “by-hand”. The utility of this machine learning approach
can be seen in particular by the fits of the low-energy neutral dissociation
cross section, plotted in Figure \ref{fig:Neutral-dissociation-cross}(a),
and the NDA-DEA cross section, plotted in Figure \ref{fig:Dissociative-electron-attachment}.
In both cases, the model succeeds in deriving a plausible cross section
in its entirety from the swarm data.

\section{\label{sec:Conclusion}Conclusion}

We have presented a set of electron-THF cross sections that refines
that constructed by de Urquijo \textit{et al}. \citep{DeUrquijo2019a}
by modifying its quasielastic MTCS, neutral dissociation, ionisation
and electron attachment cross sections. A unique aspect of this work
is that these proposed modifications were performed automatically
by a neural network model that was trained in order to solve the electron-THF
inverse swarm problem for realistic sets of cross sections taken from
the LXCat project \citep{Pancheshnyi2012,Pitchford2017,LXCat}. The
resulting set of THF cross sections was found to be self-consistent,
in that it accurately reproduced many of the swarm measurements that
were used to perform the fit. It was thus concluded that the resulting
machine-refined cross section set was of a comparable quality to the
hand-refined set of de Urquijo \textit{et al}. \citep{DeUrquijo2019a},
though it was noted that both sets have their own strengths and weaknesses.
Taking the subjectivity out of forming recommended cross section data
sets (i.e. the ``by-hand'' approach adopted previously in Ref. \citep{DeUrquijo2019a}),
for describing the behaviour of electrons as they travel through a
background gas under the influence of an applied external electric
field, is an important development and while further work clearly
needs to be undertaken on our current neural network approach this
study represents a step forward in achieving that goal.

Of the modifications to the de Urquijo \textit{et al}. \citep{DeUrquijo2019a}
set that were proposed by the neural network, the largest changes
were made to the low-energy neutral dissociation cross section, plotted
in Figure \ref{fig:Neutral-dissociation-cross}(a), and the electron
attachment cross section, plotted in Figure \ref{fig:Dissociative-electron-attachment}.
This was expected, as no explicit constraints were placed on these
cross sections in Figure \ref{fig:Confidence-band-of}, leaving the
neural network with the task of determining both in their entirety
using the swarm data alone. This task of simultaneously determining
multiple unknown cross sections entirely from swarm data is a daunting
prospect and the apparent success of the neural network in this case
highlights the utility of this automated approach to swarm analysis.

One limitation of the specific machine learning approach taken here
is that it provides only a single proposed THF cross section set when
it is evident that multiple are plausible. We intend to address this
non-uniqueness of the inverse swarm problem through the use of alternative
neural network architectures that allow for the uncertainty in the
predicted cross sections to be quantified. Examples of such alternatives
include mixture density networks \citep{Bishop1994} and conditional
generative models \citep{Sohn2015,Mirza2014,Dinh2014,Dinh2016,Kingma2018}.

In the future, we plan to also apply machine-assisted swarm analysis
toward determining cross sections for other important molecules of
biological interest, including tetrahydrofurfuryl alcohol (THFA) \citep{Jones2013,Limao-Vieira2014,Duque2014a,Mozejko2006,Bellm2012,Duque2014,Chiari2014,Brunger2017}
and water \citep{White2014}. It is promising to note that, due to
the data-driven nature of machine learning, such machine-adjusted
cross section sets can continue to be revisited as the LXCat databases
continue to grow and be refined \citep{Pancheshnyi2012,Pitchford2017}.

\ack{}{}

The authors gratefully acknowledge the financial support of the Australian
Research Council through the Discovery Projects Scheme (Grant \#DP180101655).
JdeU thanks PAPIIT-UNAM, Project IN118520 for support. G.G. acknowledges
support from the Spanish Ministerio de Ciencia, Innovación y Universidades-MICIU
(Projects FIS2016-80440 and PID2019-104727RB-C21) and CSIC (Project
LINKA 20085).

\appendix

\bibliographystyle{iopart-num}
\addcontentsline{toc}{section}{\refname}\bibliography{references}

\end{document}